\documentclass[twocolumn,preprintnumbers,amsmath,amssymb]{revtex4}
\usepackage{graphicx}
\usepackage{dcolumn}
\usepackage{bm}

\usepackage{graphicx}  
\begin{document}

\title{
Effect of self-consistent electric field on characteristics of
graphene p-i-n tunneling transit-time diodes
}
\author{
V. L. Semenenko$^1$, V.~G. Leiman$^1$, A.~V.~Arsenin$^1$, 
V.~Mitin$^2$,  M. Ryzhii$^{3,5}$, T.~Otsuji$^{4,5}$, and
V~Ryzhii\footnote{Electronic mail: v-ryzhii(at)riec.tohoku.ac.jp}$^{4,5}$,
}
\affiliation{
$^1$ Department of General Physics, 
Moscow Institute of Physics and Technology,
Dolgoprudny, Moscow Region 141700, Russia\\
$^2$ Department of Electrical Engineering, University at Buffalo, Buffalo,
NY 1460-1920, U.S.A.\\
$^3$Computational Nanoelectronics Laboratory, University of Aizu, Aizu-Wakamatsu  965-8580, Japan\\
$^4$Research Institute for Electrical Communication,Tohoku University, Sendai 980-8577, Japan\\
$^{5}$Japan Science and Technology Agency, CREST, Tokyo 107-0075, Japan\\
}

\begin{abstract}
We develop a device model for p-i-n tunneling transit-time diodes based on single- and multiple graphene layer  structures operating at the reverse bias voltages. The model of the graphene tunneling transit-time diode (GTUNNETT) accounts for  the features of the  interband tunneling generation of electrons and holes and their ballistic transport
in the device i-section, as well as the effect of the self-consistent electric field
associated with the charges of propagating electrons and holes.
Using the developed model, we calculate the dc current-voltage characteristics
and the small-signal ac frequency-dependent admittance as functions of the GTUNNETT structural parameters , in particular, the number of graphene layers and the dielectric constant
of the surrounding media. It is shown
that the admittance real part can be negative in a certain frequency range.
As revealed,  if the i-section somewhat shorter than one micrometer,
 this range corresponds to the terahertz frequencies.
Due to the effect of the self-consistent electric field,
the behavior of the GTUNNETT admittance in the range 
of its negativity of  its real part 
is rather sensitive to the  relation between the number of graphene layers and dielectric constant.  The obtained results demonstrate that GTUNNETTs with optimized structure can be used in efficient terahertz oscillators.
\end{abstract}

\maketitle
\newpage
\section{Introduction}

Pioneering papers by Shur and Eastman~\cite{1,2} have stimulated
extensive studies (which continue already for the fourth decade) of ballistic
 electron and hole transport (BET and BHT, respectively), i.e., collision free transport in short semiconductor structures. The main incentive is the realization of fastest velocities of electrons/holes and, hence, achievement
of the operation of diodes and  transistors in terahertz range (THz)
of frequencies and  low power consumption. Even at the initial  stage of the ballistic transport research, several concepts of ballistic THz sources 
have been put forward (see, for instance, an early review~\cite{3}).
However, the realization THz generation in different semiconductor devices
associated with BET/BHT, in particular, analogous to vacuum devices,
meets the problems associated with electron scattering in real
semiconductor structures. Creation of heterostructures with selective
doping with a two-dimensional electron gas (2DEG) spatially
separated from the donors, has resulted in achievement of very 
long mean free path of electrons, at least at low temperatures.
Recent discoveries of unique properties of graphene~\cite{4,5},
in particular, the demonstration of possibility of very long electron and hole mean free path in graphene layers (GLs) and what is even more interesting
in multiple graphene layers (MGLs)~\cite{6,7} add optimism
in  building  graphene based THz devices using BET.
The concept of graphene tunneling transit-time (GTUNNETT) p-i-n diode , which
exhibits  a negative dynamic conductivity in the THz range, was 
proposed and substantiated in Refs.~\cite{8,9}. This concept based not only on BET or quasi-BET (as well as BHT or quasi-BHT)
in GLs and MGLs, but also on a strong interband 
tunneling under the electric field with a
pronounced anisotropy~\cite{10,11}
due to the gapless energy spectrum, 
and constant absolute value  of electrons and holes  velocities~\cite{4}.
Due to this, the electrons in the conduction band and the holes
in the valence band generated owing to the interband tunneling in the electric field
propogate primarily in the electric field direction with the velocity
(in this direction) virtually equal to the characteristic velocity
$v_W \simeq 10^8$~cm/s. A large value of the directed velocity 
in GLs and MGLs promotes the  device operation at elevated frequencies.
  
As shown~\cite{8,9}, for the self-excitation of THz oscillations in
a circuit with GTUNNETT diode, this circuit should serve as a resonator.
However, at elevated tunneling currents
in GTUNNETTs 
considered previously~\cite{8,9}
the self-consistent charge associated with propagating electron and hole streams 
can affect the spatial distribution of the self-consistent electric field
and the electric potential in the i-section.
As a result, the self-consistent electric field
near the p-i- and i-n-juctions can  be substantially reinforced.
This, in turn,  influences
 the tunneling generation of electrons and holes
and
the their transit conditions and, hence, the GTUNNETT dc and ac characteristics.

In this paper, in contrast to the previous treatment~\cite{8,9}, we account for  the self-consistent electric field associated with the
variations of the electron and hole lateral charges in the i-section and their effect on
the injection and the dc and ac characteristics. The effects of the space charge in planar
TUNNETTs with the propagation of carriers perpendicular to the structure plane were considered
by Gribnikov et al.~\cite{12}.
The problems of calculation of the dc and ac characteristics of devices based on lateral structures
accounting for the self-consistent electric field are substantially complicated by
the features of the structure geometry (2D electron and hole channels and blade-like contact regions).
In particular, as shown in the following, the related mathematical problems 
are reduced to a system of rather complex nonlinear integral-differential
 equations.
Using the GTUNNETT device model, we derive these equations, solve them numerically, and find the characteristics.

\begin{figure}[t]\label{Fig.1}
\begin{center}
\includegraphics[width=6.3cm]{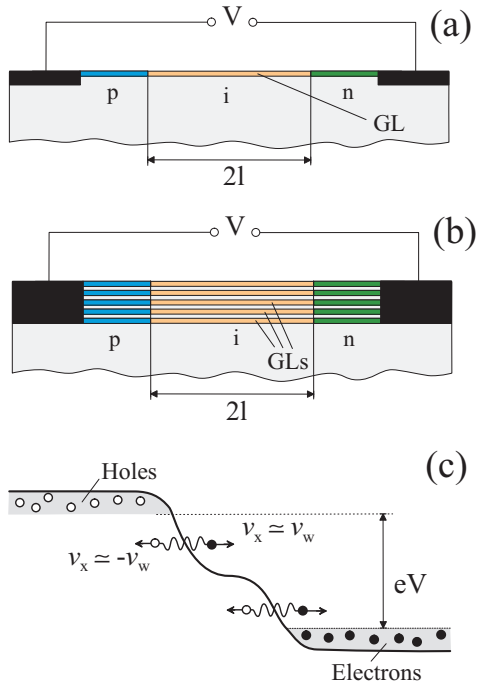}
\caption{
Schematic views of  GTUNNETT p-i-n diodes  (a) with a single GL, (b) with an MGL
structure, 
and (s) their band diagram  at reverse bias. Arrows show the propagation directions  of 
electrons and holes generated due to interband tunneling (mainly in those regions, where the electric field is relatively strong).}
\end{center} 
\end{figure} 

\section{Equations of the model}

The device under consideration is based on a GL or MGL structures
with p- and n- side sections with the contacts and an undoped
transit i-section.
For definiteness, we assume that p-and  n-sections in a GL
or in an MGL are created by chemical doping~\cite{13},
although in similar devices but with extra gates
these sections can be formed  electrically~\cite{10,11}. 
Under the reverse bias voltage $V_0$, the potential 
drops primarily across the i-section 
A schematic view of the structures in question and their band-diagram
 under sufficiently strong  reverse bias ($V_0 > 0$) corresponding to the potential distribution in the i-section 
 with a markedly nonuniform electric field
are shown in Fig.~1.

The electrons and holes injected into the i-section of one GL 
are characterized by
the electron and hole sheet concentrations, $\Sigma^- = \Sigma^-(t,x)$ and $\Sigma^+ = \Sigma^+(t,x)$, respectively
and the potential $\varphi = \varphi (t, x, y)$. Here the axis $x$
is directed along the GL (or MGL) plane, i.e., in the direction of the current, the axis $y$ is directed perpendicular 
to this plane, 
and $t$ is the time. In the  general case when the i-section is based on MGL, 
$\Sigma^- = \Sigma^-(t,x)$ and $\Sigma^+ = \Sigma^+(t,x)$ are the  electron and hole
densities in each GLs. In the case of BET and BHT, on which we focus mainly,
the electron and hole sheet densities in each GL obey the continuity equations  

\begin{equation}\label{eq1}
\frac{\partial\Sigma^{\mp}}{\partial t} \pm v_W\frac{\partial\Sigma^{\mp}}{dx} = g,
\end{equation}
and the Poisson equation

\begin{equation}\label{eq2}
\frac{\partial^2\Phi}{\partial x^2} 
+ \frac{\partial^2\Phi}{\partial y^2} 
= \frac{4\pi\,eK}{\ae}(\Sigma^- - \Sigma^+)\cdot\delta(y),
\end{equation}
respectively.
Here $e = |e|$ is the electron charge, $\ae$ is the dielectric constant of the media 
surrounding GL (or MGL structure), $K$ is the number of GLs in the GTUNNETT structure, and $\delta(y)$ is the delta function reflecting a narrowiness of  GL and 
 MGL structures even with rather large number of GLs
in the $y-$ direction.
Equation (1) corresponds to the situation when the electrons and holes generated due to the interband tunneling in the i-section
obtain the directed velocities $v_x = v_W$ and $v_x =- v_W$ and preserve them during the propagation. 
The boundary conditions correspond to the assumption that the electrons and holes appear
in the i-section only due to the interband tunneling (the injection of electrons from the p-section and holes from
the n-section is negligible) and that the highly conducting side contacts to the p- and n-sections are
of blade type (the thicknesses of GL and MGL and the contacts to them
are much smaller than the spacing between the contacts):

\begin{equation}\label{eq3}
\Sigma^{-}|_{x = -l} = 0, \qquad \Sigma^{+}|_{x = -+l} = 0,
\end{equation}

\begin{equation}\label{eq4}
\Phi|_{x \leq  -l, y = 0} = -V/2, \qquad \Phi|_{x \geq +l, y = 0} = V/2,
\end{equation}
where $2l$ is the length of the i-section and $V = V_0 + \delta V_{\omega}\exp(-i\omega t)$
is the net voltage, which comprises the bias voltage $V_0$ and the signal component with the amplitude $\delta V$
and the frequency $\omega$.
The interband tunneling generation rate of electrons and holes in each GL (per unit of its area)
is given by~\cite{8,9,10,11}

\begin{equation}\label{eq5}
g = 
g_0\biggl(\frac{2l|\partial \varphi/\partial x|}{V_0}\biggr)^{3/2},
\end{equation}
where  $\varphi(x) = \Phi(x,y)|_{ y = 0}$, $\hbar$ is the Planck constant,
$$ 
g_0 = \frac{v_W}{2\pi^2}\biggl(\frac{eV_0}{2l\hbar\,v_W}\biggr)^{3/2},
$$
so that the characteristic tunneling dc current (per unit length in the transverse
$z$-direction) and the characteristic electron and hole sheet density can be presented as

$$
J_{00} = 4elg_0, \qquad \Sigma_0 = 2lg_0/v_W.
$$

From equation~(2) with boundary condition~(4), for the potential $\varphi(x)$ in the GL (or MGL structure) plane
we obtain
$$
\varphi (x) = \frac{V}{\pi}\sin^{-1}\biggl(\frac{x}{l}\biggr) 
$$
\begin{equation}\label{eq6}
+
\frac{Ke}{\ae}\int_l^l[\Sigma^+(x^{\prime}) - \Sigma^-(x^{\prime})]G\biggl(\frac{x}{l},\frac{x^{\prime}}{l}\biggr)dx^{\prime}
\end{equation}
where

\begin{equation}\label{eq7}
G\biggl(\frac{x}{l},\frac{x^{\prime}}{l}\biggr)
 = \frac{1}{2\pi}\biggl|\frac{\sin[(\cos^{-1}\xi + \cos^{-1}\xi^{\prime})/2]}
{\sin[(\cos^{-1}\xi - \cos^{-1}\xi^{\prime})/2]}\biggr|.
\end{equation}

Introducing the dimensionless quantities:
$n^{\mp} = \Sigma^{\mp}/\Sigma_0$,  $\xi = x/l$,  
$\xi^{\prime} = x^{\prime}/l$, and $\tau = v_Wt/l$, 
Eqs. (1) and (6) can be reduced to the following system of nonlinear
integro-differential equations:

$$
\frac{\partial n^{\mp}}{\partial \tau} \pm \frac{\partial n^{\mp}}{\partial\xi}
= \biggl[\frac{V/V_0}{\pi\sqrt{1 - \xi^2}}
$$
\begin{equation}\label{eq8}
+ \gamma\int_{-1}^{1}[n^+(\xi^{\prime},\tau) - n^-(\xi^{\prime},\tau)]
\frac{\partial}{\partial\xi}G(\xi,\xi^{\prime})\,d\xi^{\prime}
\biggr]^{3/2}.
\end{equation}
Here
\begin{equation}\label{eq9}
\gamma = \frac{2Kel\Sigma_0}{\ae\,V_0} = \frac{Ke^2}{4\pi^2\ae\hbar\,v_W}\sqrt{\frac{2leV_0}{\hbar\,v_W}}.
\end{equation}
\begin{equation}\label{eq10}
\frac{\partial}{\partial\xi}G(\xi,\xi^{\prime}) = \frac{1}{\pi(\xi^{\prime} - \xi)}\frac{\sqrt{1 - \xi^{\prime \,2}}}{\sqrt{1 - \xi^2}}.
\end{equation}
The term with the factor $\gamma$ in the right-hand side of Eq.~(8) is associated with the contributions of the electron and hole charges to the self-consistent electric field in the i-section.

The dc and ac components of the terminal dc and ac currents, 
$J_0$ and $\delta J_{\omega}$, are expressed via the dc and ac components 
$n^{\mp}_0$ and $\delta n^{\mp}_{\omega}$ as follows:
\begin{equation}\label{eq11}
J_0 = ev_WK\Sigma_0(n_0^{+} + n_0^{-}),
\end{equation}

\begin{equation}\label{eq12}
\delta J_{\omega} = ev_WK\Sigma_0\int_{-1}^1\rho(\xi)(\delta n_{\omega}^{+} + \delta n_{\omega}^{-})d\xi - i\omega\,C\delta V_{\omega}.
\end{equation}
Here $\rho(\xi) = 1/\pi\sqrt{1 - \xi^2}$ is the form factor and $C \sim \ae/2\pi^2$ is the geometrical capacitance~\cite{14,15}.
The explicit coordinate dependence of the form factor is a consequence of the Shocley-Ramo theorem~\cite{16,17} for the device geometry under consideration.

Hence the GTUNNETT small- signal admittance $Y_{\omega} = \delta J_{\omega}/\delta V_{\omega}$
is presented in the form
\begin{equation}\label{eq13}
Y_{\omega} = ev_WK\Sigma_0\int_{-1}^1\rho(\xi)\frac{d}{d\delta V_{\omega}}(\delta n_{\omega}^{+} + \delta n_{\omega}^{-})d\xi - i\omega\,C.
\end{equation}

As follows from Eq. (8), the problem 
under consideration is characterized by the parameter $\gamma$.

If $\ae = 1.0$, $K = 1$, $2l =0.7~\mu$m, in the voltage range  $V_0 = 100 - 200$~mV 
one obtains $\Sigma_0 \simeq (0.6 - 1.7)\times 10^{10}$~cm$^{-2}$ and 
$\gamma  \simeq 0.61 - 0.87$.

\section{Spatial potential  distributions and current-voltage characteristics}
If the  charges created by the propagating electrons and holes in the i-section are insignificant, that corresponds to $\gamma \ll 1$. In this case,
the potential distribution is given by

\begin{equation}\label{eq14}
\varphi_0 (x) \simeq \frac{V_0}{\pi}\sin^{-1}\biggl(\frac{x}{l}\biggr),
\end{equation}
and 
Eqs. (8), neglecting the term with $\gamma$,  can be solved analytically. Taking into account boundary conditions~ Eq.~(3), from Eqs. (8) we obtain
\begin{equation}\label{eq15}
n_0^+ + n_0^- = \frac{1}{\pi^{3/2}}\int_{-1}^{1}\frac{d\xi}{(1 - \xi^2)^{3/4}}
= const.
\end{equation}
After that,
using Eq. (11) and considering Eq.~(15), one can find the following formula
for the dc current-voltage characteristic:
$$
J_0 = 
\frac{KJ_{00}}{\pi^{3/2}}\int_0^1\frac{d\xi}{(1 - \xi^2)^{3/4}}
$$
\begin{equation}\label{eq16}
= K\frac{\Gamma(1/4)\Gamma(1/2)}{\Gamma(3/4)}\frac{ev_W}{\pi^{7/2}2\sqrt{2l}}\biggl(\frac{eV_0}{\hbar\,v_W}\biggr)^{3/2},
\end{equation}
where $\Gamma(x)$ is the Gamma-function.
A distinction between $J_0$ and $J_{00}$ is due to the nonuniformity of the
electric field in the i-section associated with the feature of the device geometry
taken into account calculating $J_0$.
At $K = 1$, $2l = 0.7~\mu$m and $V_0 = 100 - 200$~mV, Eq.~(16) yields
$J_0 \simeq 0.18 - 0.51$~A/cm.

\begin{figure}[t]\label{Fig.2}
\begin{center}
\includegraphics[width=6.5cm]{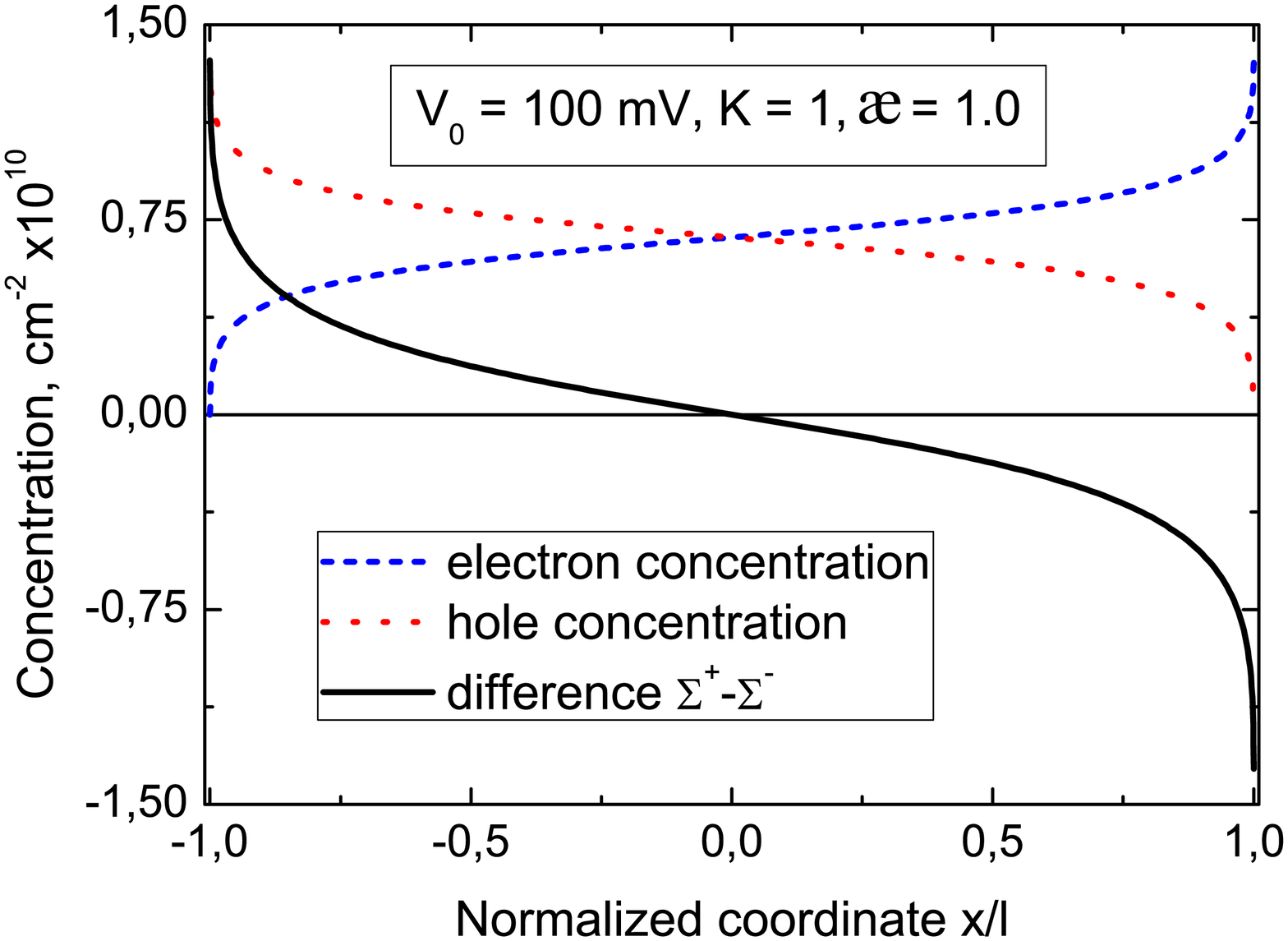}\\
\includegraphics[width=6.5cm]{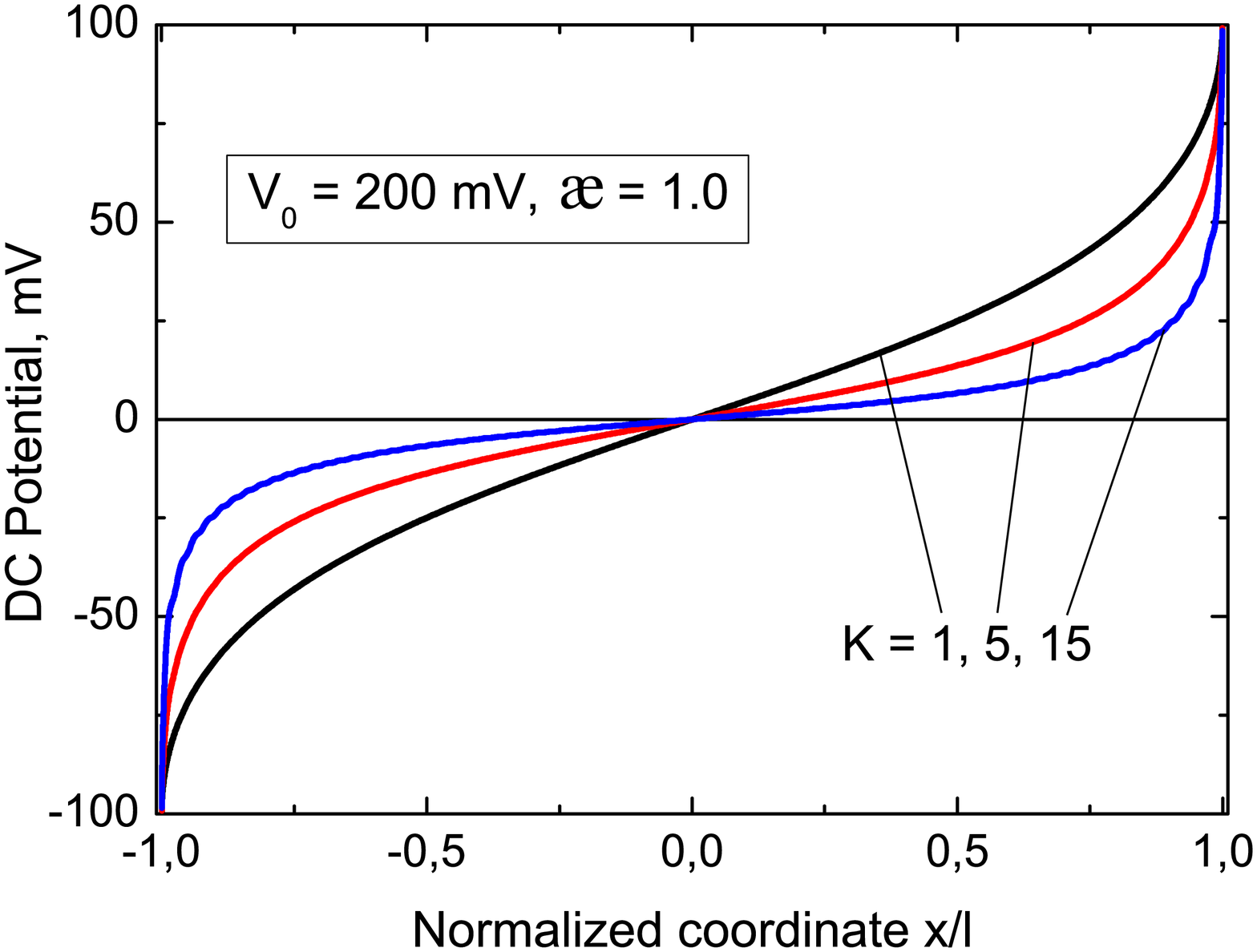}
\caption{Spatial distributions of electron and hole concentrations
(upper panel) and 
of electric potential 
(lower panel) in the GTUNNETT i-section.}
\end{center} 
\end{figure} 
\begin{figure}[t]\label{Fig.3}
\begin{center}
\includegraphics[width=6.5cm]{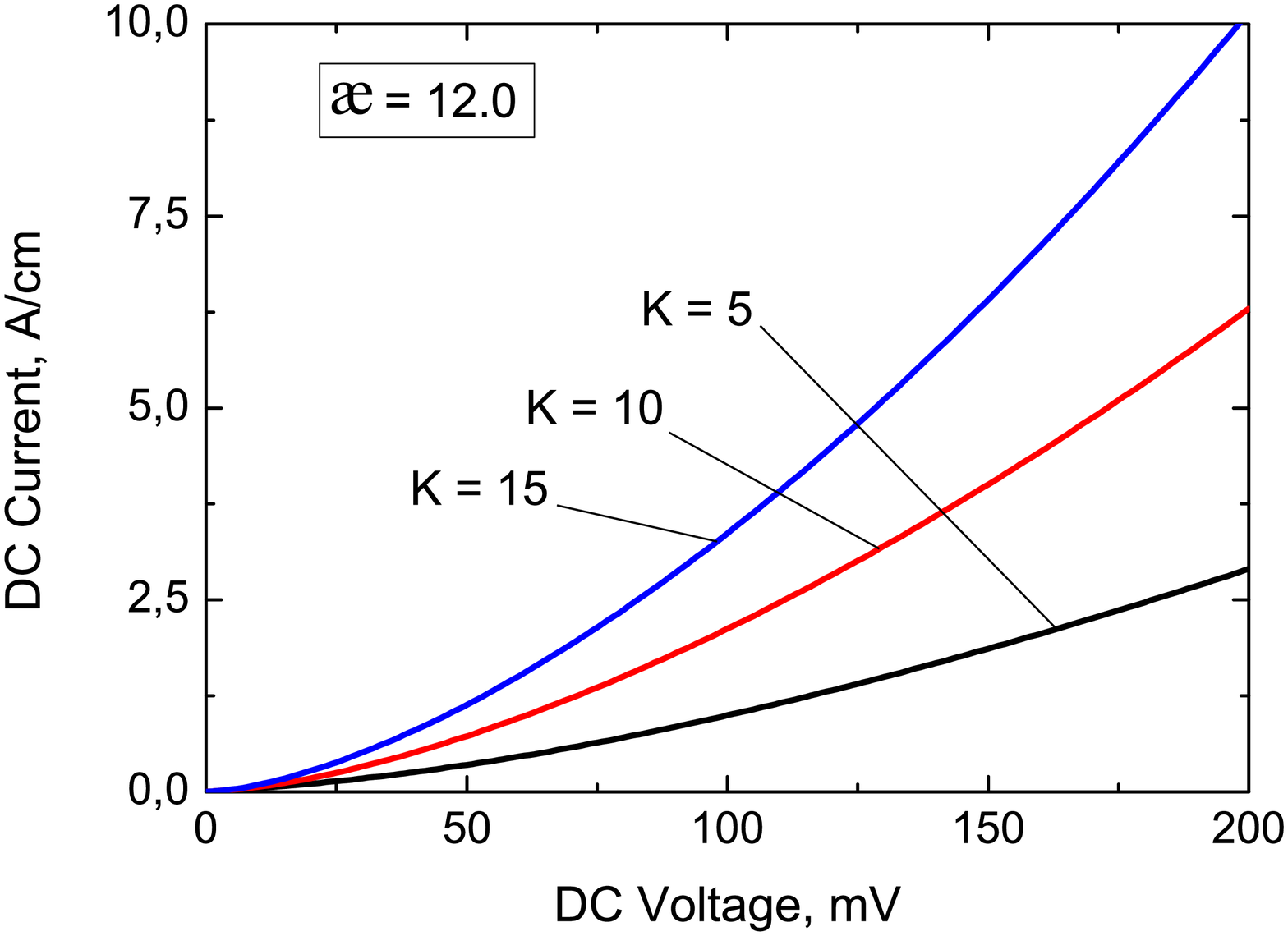}\\
\caption{Current-voltage characteristics for GTUNNETTs with different 
number of GLs $K$.
}
\end{center} 
\end{figure} 

\begin{figure}[t]\label{Fig.4}
\begin{center}
\includegraphics[width=6.5cm]{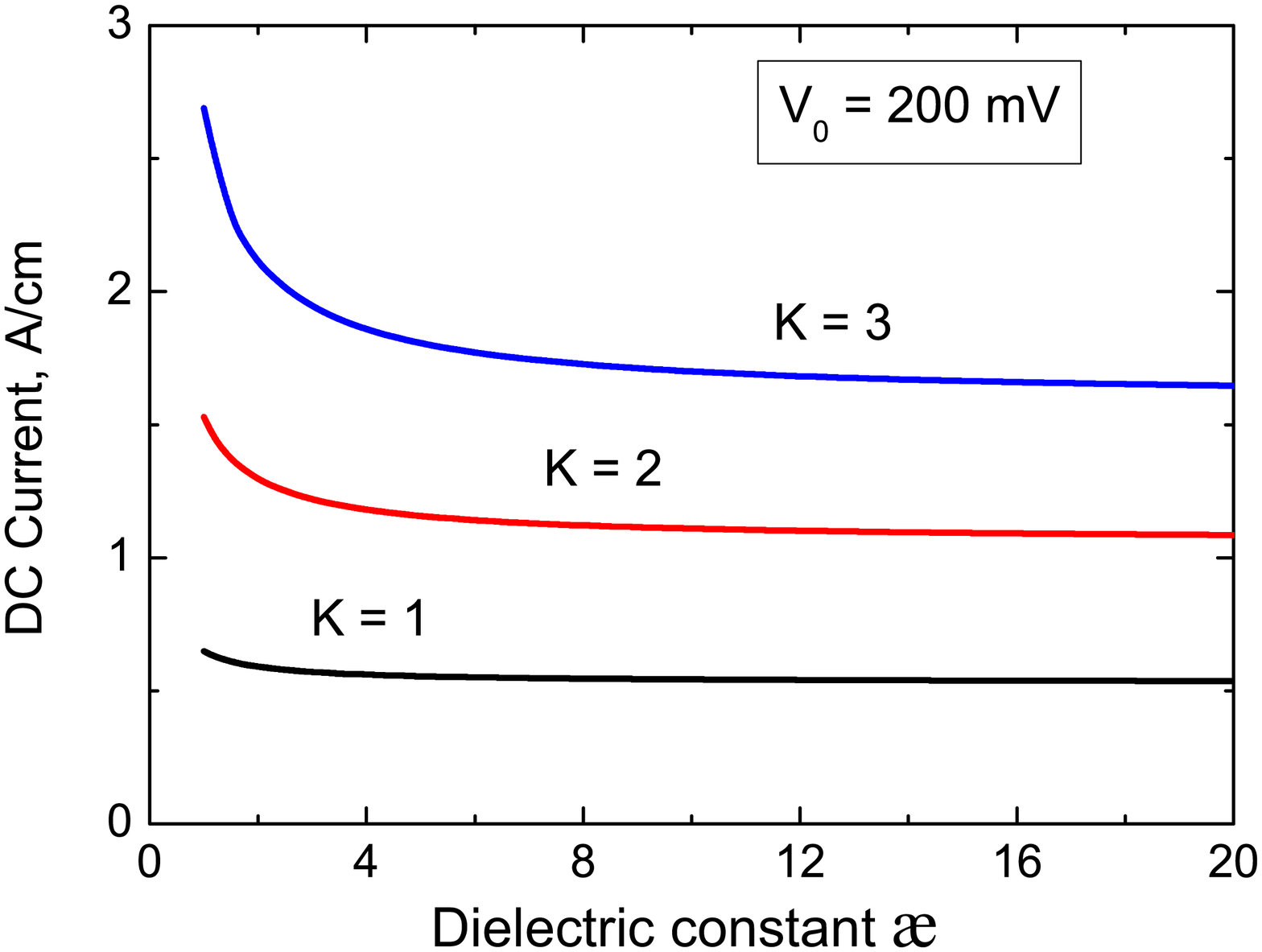}\\
\includegraphics[width=6.5cm]{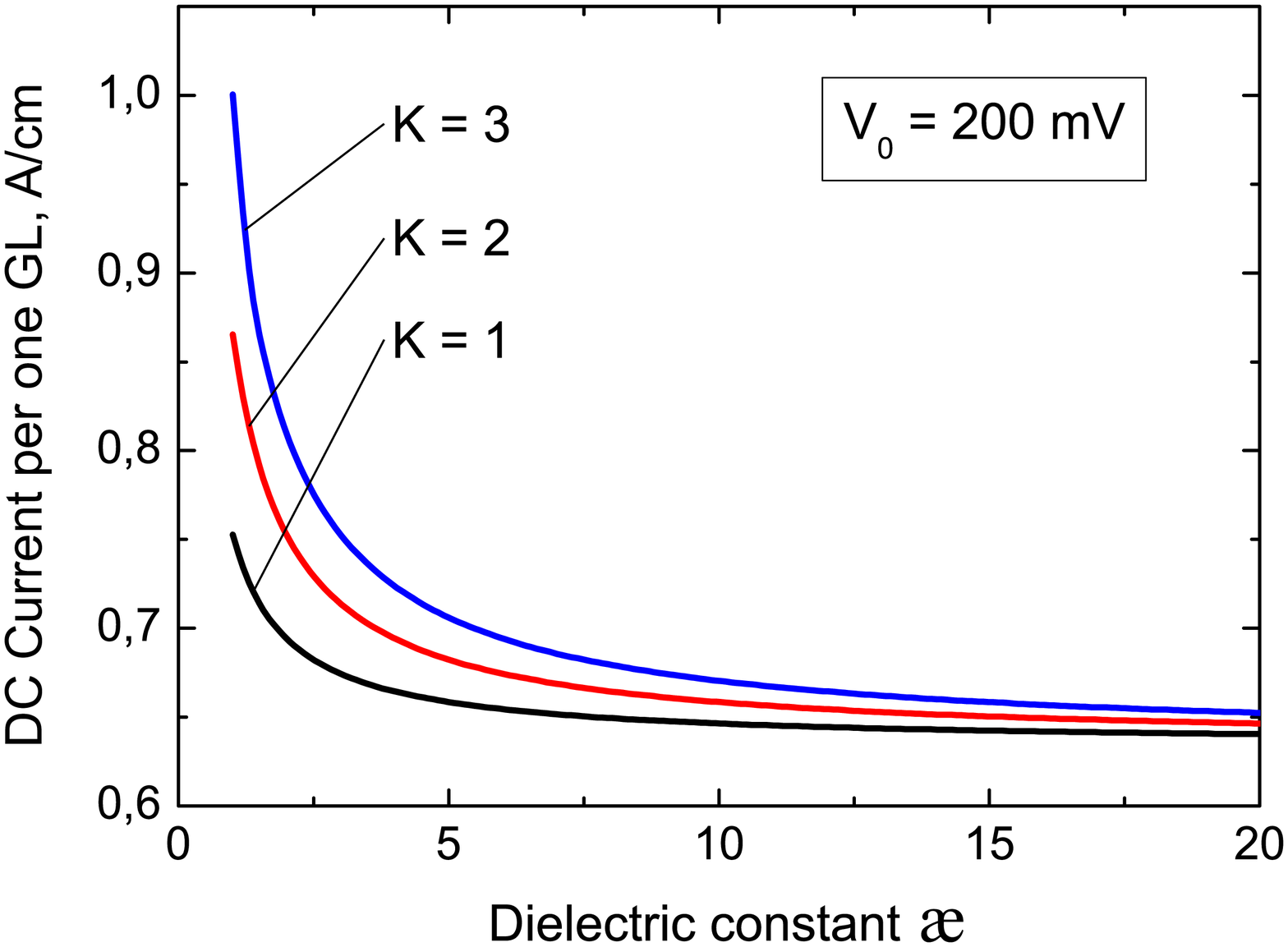}
\caption{Dependences of dc current on dielectric constant  
for different numbers of GLs $K$: net current - upper panel and current per one GL - lower panel.}
\end{center} 
\end{figure} 

\begin{figure}[t]\label{Fig.5}
\begin{center}
\includegraphics[width=6.5cm]{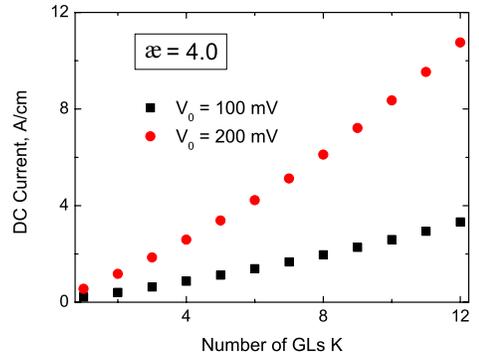}\\
\caption{Dependences of dc current on number of GLs  $K$
at  different voltages.}
\end{center} 
\end{figure}

To take the effect of the  self-consistent electric field on the tunneling, one needs to solve
system of Eqs.~(8). Due to a complexity of the nonlinear integro-differential 
equations in question, a numerical approach is indispensable.
Equations~(8) were solved numerically using successive approximation method,
which is valid when $\gamma < \gamma_c = 5$. In the cases $\gamma > 5$, 
the method of the parameter evolution was implemented.


 Using Eqs.~(8) and (11) and setting $V = V_0$,  we calculated numerically the GTUNNETT dc characteristics:
 spatial distributions of the dc electric potential and the dc components of the
 electron and hole densities, as well as the dc current-voltage characteristics.
 The pertinent results are shown in Figs.~2-5. 

Upper panel in Fig.~2  shows examples of the  spatial distributions 
of the electron and hole sheet concentrations in the i-section.
The spatial distributions 
of the dc electric
potential  calculated
 for GTUNNETTs with different numbers of GLs $K$
at fixed voltage are shown in Fig.~2 (lower) panel. One can see that an increase in $K$ results
in a marked concentration of the electric field near the doped sections. This is because at larger $K$, the net tunneling generation rate becomes stronger. This, in turn, results in higher 
 charges of propagating electron and hole components, particularly,
 near the p-i- and i-n-junctions, respectively.

Figure 3 demonstrates a difference in the dc current-voltage characteristics in GTUNNETTs 
with different numbers of GLs $K$.
An increase in $K$ leads to an increase in the dc current not only because of the increase in the number of current channels but because of the reinforcement of the self-consistent electric field near
the edges of the i-section(as mentioned above).
As can be seen in Fig.~4  (upper panel), the three-fold  increase in  $K$ 
leads to more than three-fold increase in the dc current particularly  at relatively small dielectric constants. This is confirmed by  plots in Fig.~4 (lower panel),  from which  it follows that the dc current in each GL 
is larger in the devices with larger   number of GLs.  It is worth noting that when the dielectric constant is sufficiently large (about 15-20), the dc current in one GL is virtually the same in the devices with different number of GLs,
because in this case the charge effect is suppressed. 
In this case the value of the  dc current
(per one GL) is approximately the same as that obtained in the above numerical estimate using analytical formula given by Eq.~(16).
In GTUNNETTs with a moderate value of $\ae$, the charge effect leads to
superlinear dependences of the net dc current vs number of GLs
 as seen in Fig.~5.

\section{GTUNNETT admittance}

\begin{figure}[t]\label{Fig.6}
\begin{center}
\includegraphics[width=6.5cm]{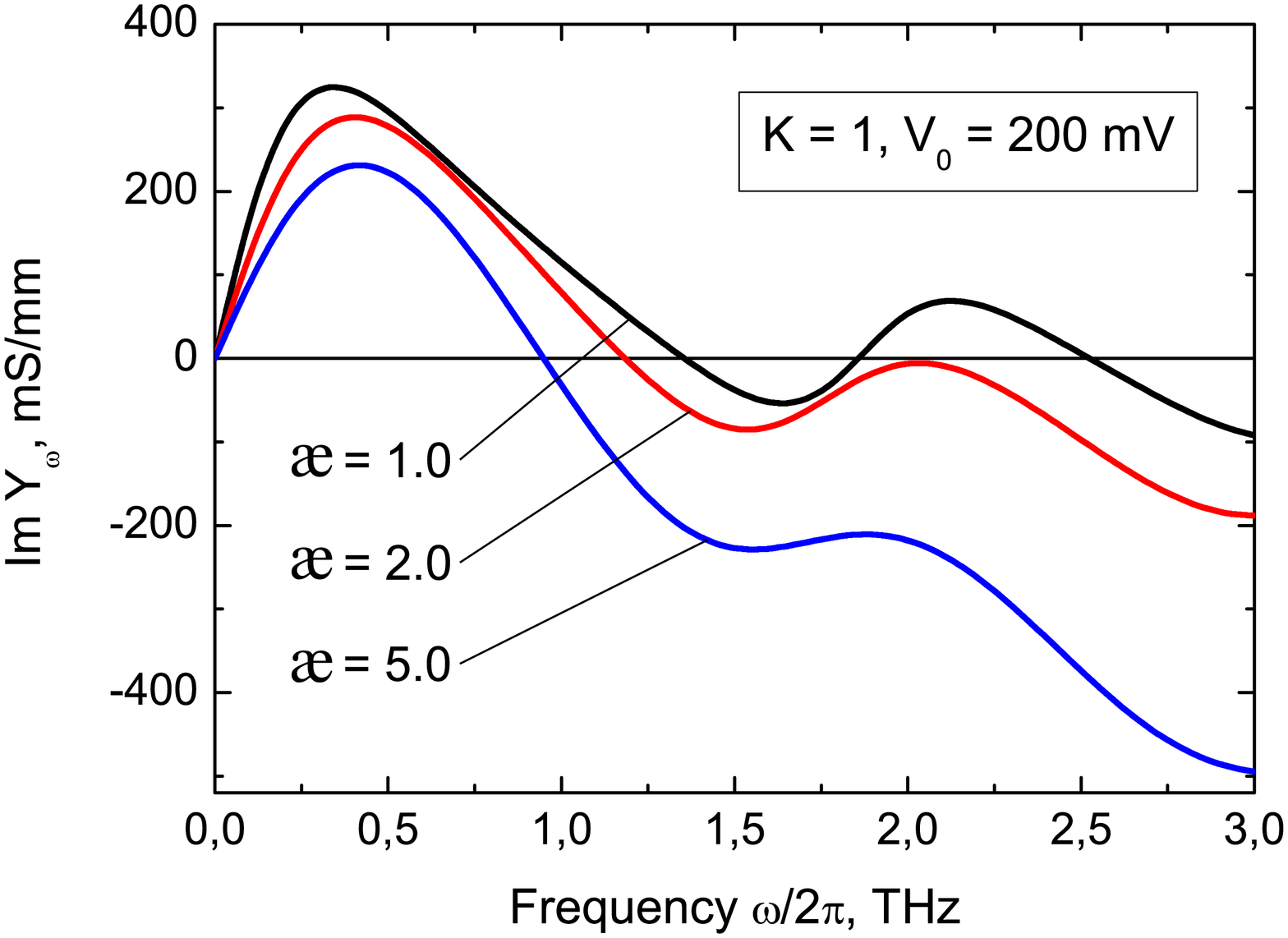}\\
\includegraphics[width=6.5cm]{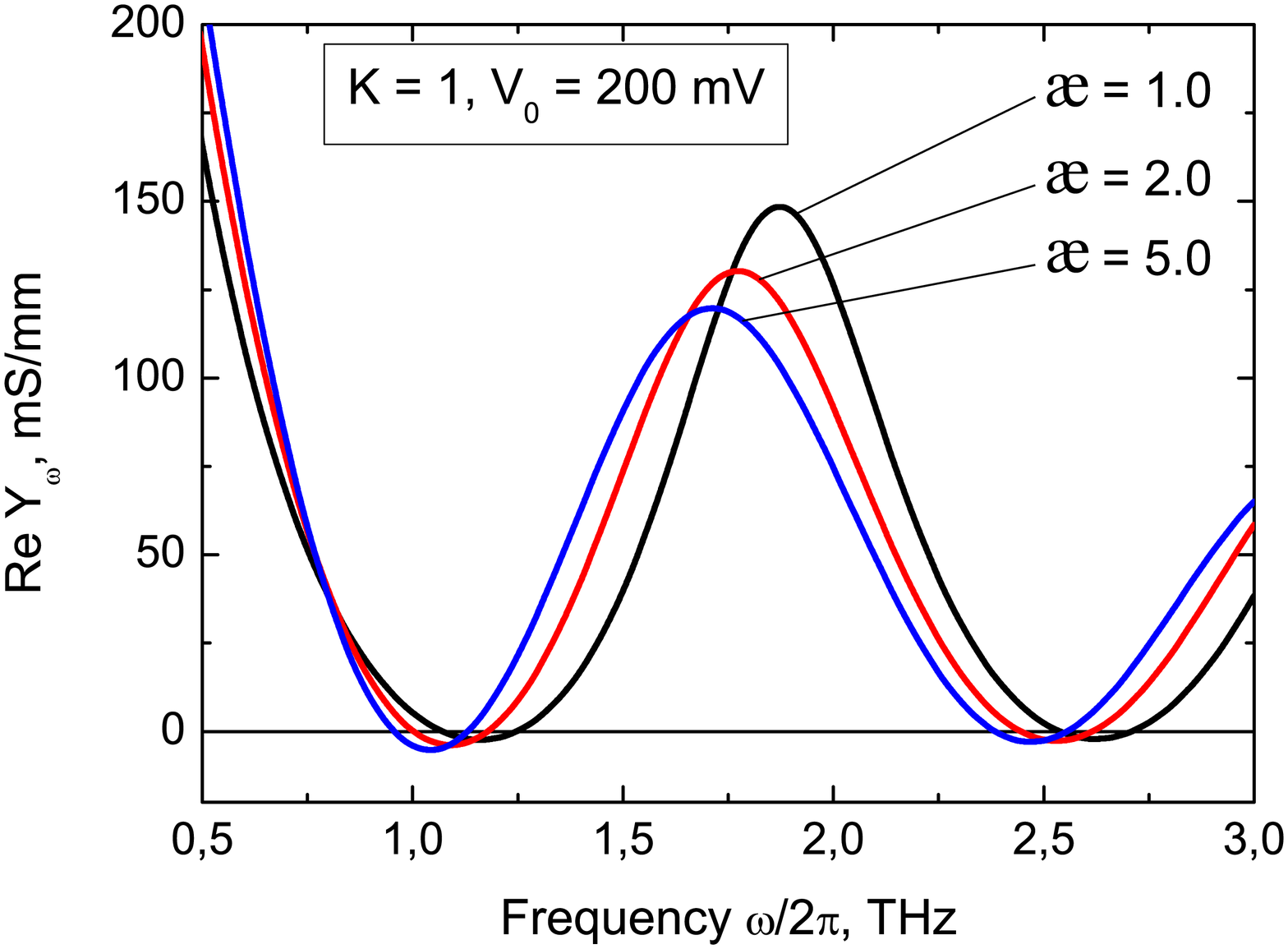}
\caption{
Frequency dependences of imaginary  part, Im~$Y_{\omega}$,   and real  part, Re~$Y_{\omega}$, of GTUNNETT admittance 
(upper and lower panels, respectively)
at different values
of dielectric constant $\ae$ ($K = 1$ and $2l = 0.7~\mu$m). }
\end{center} 
\end{figure} 
\begin{figure}[t]\label{Fig.7}
\begin{center}
\includegraphics[width=6.5cm]{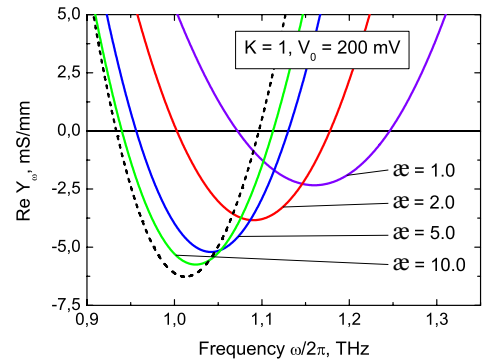}
\caption{
Frequency dependences of real  part, Re~$Y_{\omega}$,  of GTUNNETT admittance  in the frequency range,
 where Re~$Y_{\omega} < 0$
at different values
of dielectric constant $\ae$ ($K = 1$ and $2l = 0.7~\mu$m) calculated accounting for electron and hole charges (solid lines) and neglecting them (dashed line)~\cite{9}.}
\end{center} 
\end{figure}

\begin{figure}[t]\label{Fig.8}
\vspace*{-0.4cm}
\begin{center}
\includegraphics[width=6.5cm]{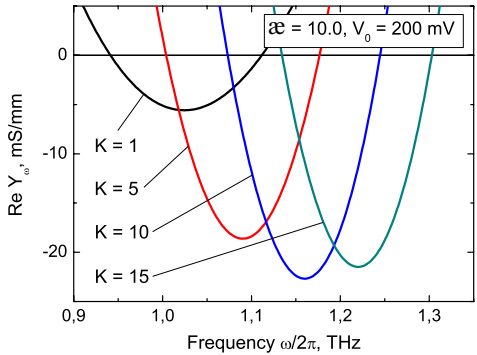}
\caption{
Frequency dependences of  real part of admittance Re~$Y_{\omega}$
for GTUNNETTs with different number of GLs K
($\ae = 10.0$  and $2l = 0.7~\mu$m).}
\end{center} 
\end{figure}
\begin{figure}[t]\label{Fig.9}
\begin{center}
\includegraphics[width=6.5cm]{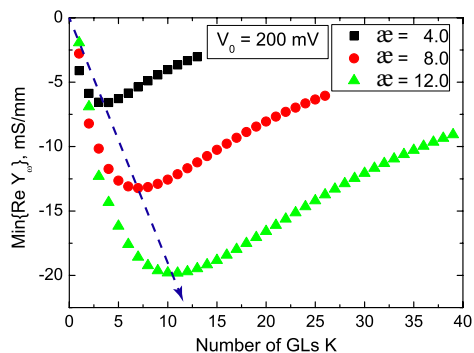}
\caption{
Minimum value of real part of GTUNNETT admittance as a function of number of GLs  $K$
for GTUNNETTs with different dielectric constant $\ae$.}
\end{center} 
\end{figure}

Disregarding the term in Eq.~(8) associated with the electron and hole charges in the i-section, and using Eq.~(12) , one can obtain analytically the following formulas for the ac potential
distribution in the i-section, and the imaginary and real parts of the GTUNNETT admittance ~\cite{9}:
\begin{equation}\label{eq17}
\delta\varphi (x)_{\omega} = \frac{\delta\,V_{\omega}}{\pi}\sin^{-1}\biggl(\frac{x}{l}\biggr) ,
\end{equation}
\begin{equation}\label{eq18}
{\rm Im} Y_{\omega} =  Y_0 \sin(\omega\tau_t)\cdot {\cal J}_0(\omega\tau_t) -  \omega C,
\end{equation}
\begin{equation}\label{eq19}
{\rm Re} Y_{\omega} = Y_0 \cos(\omega\tau_t)\cdot {\cal J}_0(\omega\tau_t),
\end{equation}
where $Y_0 = dJ_0/dV_0 = 3J_0/2V_0 \propto \sqrt{V_0}$ is the dc differential conductivity, $C$ is the geometrical capacitance, $\tau_t = l/v_W$ is the characteristic transit time of electrons and holes across the i-section,  and 
${\cal J}_0 (\omega\tau)$ is the Bessel function. As follows from Eqs.~(18) and (19), the admittance imaginary and real 
parts oscillate as functions of the transit angle $\omega\tau_t$ with
Re~$Y_{\omega} < 0$  at $\omega\tau_t$ near the transit time resonances
$\omega\tau_t = (2n - 1/2)\pi$, where $n = 1,2,3...$ is the resonance index.

The numerical solution of Eqs.~(8), accounting for the ac components 
of the applied voltage $\delta V_{\omega}$ and the the self-consistent charges and electric field,  shows that the oscillatory frequency dependences are preserved 
in this more realistic case although they are be quantitatively  modified.
In particular, these dependences vary with varying such parameters as the number of GLs $K$ and dielectric constant $\ae$. 
The results of numerical calculations using Eq.~(8)
are demonstrated in Figs.~6 - 9.

Figure 6 shows the dependence of the imaginary and real parts, Im~$Y_{\omega}$ and Re~$Y_{\omega}$,
 of the admittance on the signal frequency calculated 
 for GTUNNETTs with $2l = 0.7~\mu$m and $K = 1$ at $V_0 = 200$~mV
 at different values of  dielectric constant $\ae$. As follows from Fig.~6, the imaginary and real parts of the GTUNNETT admittance are oscillatory functions of the signal frequency [see also Eqs.~(18) and (19)].  
The oscillatory behavior is due to the transit-time resonances.
At sufficiently high frequencies,  the admittance imaginary part  is negative because it
is determined primarily by the geometrical capacitance, so that Im $Y_{\omega} \simeq -\omega\,C$. 
 In the certain frequency ranges, the admittance real part 
  is negative as shown in Fig.~6 (lower panel). These ranges correspond to the transit-time resonances.
 As  seen from Fig.~7 , the minima of the 
 admittance real part (where the latter is negative) become deeper and the minima shift toward smaller frequencies with increasing dielectric constant $\ae$. 
 This is because an increase in $\ae$ leads to weakening of the  effect of the electron and hole charges on the injection and propagation processes.
 The dashed line corresponds to the case when the role of these charges is diminished
 (very large $\ae$) considered previously~\cite{9}. 

Figure 8 shows the admittance real part vs signal frequency calculated for GTUNNETTs with different number of GLs $K$. One can see that GTUNNETTs with larger number of GLs $K$ exhibit much more deep minima (compare the curves which correspond to
$K = 1$ and $\ae = 10.0$ in Fig.~7 and that for $K = 10$ and $\ae = 10.0$
in Fig.~8.). As follows from this comparison, the ten-fold increase in $K$ leads to
about four-fold increase in the minimum depth.

Figure 9 shows the dependence of the minimum value of the real part of the GTUNNETT admittance
min $\{{\rm Re} Y_{\omega}\}$   on the number of GLs $K$ 
calculated for different dielectric constants $\ae$. The nonmonotonic character of these dependences is associated with the interplay of the factors determined by the influence of electron and hole charges:  an increase in $\ae$
leads to a suppression of these charges role, while
 an increase in $K$ results in
proportional reinforcement of the effect  of  these charges accompanied by
proportional increase in the current.

\section{Discussion}

Calculating the dc current, we disregarded 
the contributions of the thermogeneration of electron-hole pairs in the i-section
and the injection of minority carriers from the n- and p-sections.

The thermogeneration dc current is given by
\begin{equation}\label{eq20}
J_0^{therm} = 4Kelg_0^{therm}.
\end{equation}
The thermogeneration rate $g_0^{therm}$ at the temperature $T = 300$~K is estimated as~\cite{18}
$g_0^{ther} \simeq 10^{20} - 10^{21}$~cm$^{-2}$s$^{-1}$. As a result, at $K = 1$
and $2l = 0.7~\mu$m from Eq. (20)one obtains
$J_0^{therm} \simeq 2 \times(10^{-3} - 10^{-2})$~A/cm.

The injection dc current can be estimated as
\begin{equation}\label{eq21}
J_0^{inj} = \frac{2KeT^2}{\pi^2\hbar^2v_w}\exp\biggl(-\frac{\mu}{T}\biggr),
\end{equation}
where $\mu$ is the Fermi energy of electrons in the n-section and holes in the p-section.
Setting $\mu = 50$~meV,  from Eq. (21) for $K = 1$ we find $J_0^{inj} \simeq 0.077$~A/cm.
Thus, $J_0^{therm}, J_0^{inj} \ll J_0$,
at least in the voltage range considered above.

The typical values of the real part of the GTUNNETT small-signal admittance in the vicinity of the first transit-time resonance ($\omega\tau_t \sim 3/2$) 
is in the range Re $Y_{\omega} \simeq - 5$ to $ - 20$~mS/mm depending on $K$ and $\ae$.
This implies that the ac resistance, $R_{\omega} = H/Y_{omega}$ of optimized
GTUNNETTs with  Re~$Y_{\omega} = - 20$~mS/mm ($K = 10$,  $\ae = 10$, and $V_0 = 200$~mV) and the width $H = (0.5 - 1)$~mm, is  $R_{\omega} \simeq 50 - 100~ \Omega$. Assuming that in this case $J_0 \simeq 0.6 - 0.8$~A/mm at $V_0 = 200$~mV,
the dc power, $P_0$, generated by the propagating electrons and holes in the contact
n- and p-regions of the device can be estimated as $P_0 \simeq 150$~mW.
When Re~$Y_{\omega} = - 5$~mS/mm ($K = 1$ and $\ae = 10$), the same resistance is provided
if $H = 2 - 4$~mm with
the dc power  $P_0 \simeq 20 - 40$~mW.

The depth of the minimum of the real part of the GTUNNETT admittance can be increased
by applying higher bias voltage. This is because the admittance is proportional
to the dc differential conductivity which, in turn, is approximately proportional to $\sqrt{V_0}$.
However, an increase in $V_0$ results in the pertinent rise of $P_0$. 
If $eV_0$ exceeds the the energy of optical phonons in GLs $\hbar\omega \simeq 200$~meV, their fast emission disrupts BET and BHT in the i-section.
The emission of optical phonon leads to some isotropization 
of the angular distribution of
the electron and hole velocities of electrons and holes.
This results in some decrease in their mean directed velocities (they become somewhat smaller than $\pm v_W$) and,
hence, in some increase in the electron and hole charges and the transit time.
Such factors promote more steep dc current-voltage characteristics and
modify the frequency dependence of the admittance. Although they should not
change the GTUNNETT performance qualitatively.

The depth of the first resonance minimum of the real part of the  GTUNNETT
admittance with the parameters used in the above estimate is of the same order of magnitude as that of the small area (about $2~\mu$m$^2$) THz resonant-tunneling diodes considered by Asada et al.~\cite{19}. 
It is worth noting that the frequency at which Im~$Y_{\omega}$ turns zero
can fall into the frequency range, where Re~$Y_{\omega} < 0$. In such a case,
GTUNNETTs can exhibit the self-excitation of THz oscillations even without
the external resonant circuit.  Moreover, the contact n- and p-sections
(both ungated as in the case of their chemical doping or gated in the case of the electrical "doping") can serve as plasma resonant cavities~\cite{9,15}. The combination of the transi-time and the plasma resonances, can substantially liberalize the self-excitation conditions and enhance the THz emission efficiency. The matching
of the transit-time and plasma resonant frequencies requires a proper choice of
the length of the n- and p-sections (from several micrometers in the ungated structures
to about one micrometer in the gated ones.  
The plasmons and  related resonances in the gated GLs were predicted and analyzed several years ago~\cite{20,21} (see also Ref.~\cite {22}. Recently, the plasmons in gated GLs were detected experimentally~\cite{23,24}.
It is interesting, that as a gate for the GL plasmon cavity  another gate can be used~\cite{25} when the contact regions constitute double-GL structures~\cite{26,27}.

\section{Conclusions}

We have developed a self-consistent device model for p-i-n GTUNNETTs
with different number of GLs
which enables the  calculation of  their realistic dc and ac characteristics.
Our calculations have shown that:\\
(i) The charges of  electrons and holes propagating in the i-section and their effect on the spatial distribution of the self-consistent electric field
increase  the steepness of the dc current-voltage characteristics. This effect is  stronger in GTUNNETTs with
larger number of GLs and can be weakened by using the media surrounding GL or MGL structure (substrates and top dielectric
layers) with elevated values of the dielectric constant;\\
(ii) The imaginary and real parts of the GTUNNETT admittance are oscillatory 
functions of the signal frequency due to the transit-time resonances.
In the certain frequency ranges, which correspond to the THz frequencies, the admittance real part can be negative
that enables the use of GTUNNETTs in the sources of THz radiation;\\
(iii) The charges of  electrons and holes  influence on the spatial distribution of the self-consistent electric field and the interband tunneling  and
can substantially affect the GTUNNETT admittance, particularly, its frequency
dependence near the transit-time resonances. The role of this effect 
enhances with increasing number of GLs in the device structure and can be markedly weakened in the devices with relatively large dielectric constant;\\
(iv) The sensitivity  of the GTUNNETT characteristics to the structural parameters
(number of GLs, dielectric constant, length of the i-section) and the bias voltage
opens up wide opportunity for the optimization of GTUNNETTs for THz oscillators.

This work was partially supported by the Russian Foundation for Basic
Research (grants 11-07-12072, 11-07-00505, 12-07-00710), by grants
from the President of the Russian Federation (Russia),
the Japan Science and Technology Agency, CREST, 
the Japan Society for Promotion of Science (Japan), and the TERANO-NSF grant (USA).

\end{document}